\documentclass[12pt]{iopart}

\usepackage{amssymb}
\usepackage{enumerate}
\usepackage{color}
\usepackage{graphicx}
\newtheorem{theorem}{Theorem}
\newtheorem{remark}{Remark}

\begin{document}

\title{Moving frames and compatibility conditions for three-dimensional director fields}

\author{Luiz C B da Silva and Efi Efrati}

\address{Department of Physics of Complex Systems,
           Weizmann Institute of Science, Rehovot 7610001, Israel}
\ead{luiz.da-silva@weizmann.ac.il}
\ead{efi.efrati@weizmann.ac.il}
\vspace{10pt}
\begin{indented}
\item \today
\end{indented}

\begin{abstract}
The geometry and topology of the region in which a director field is embedded
impose limitations on the kind of supported orientational order. These limitations manifest as compatibility conditions that relate the quantities describing the director field to the geometry of the embedding space. 
For example, in two dimensions (2D) the splay and bend fields suffice to determine a director uniquely (up to rigid motions) and must comply with one relation linear in the Gaussian curvature of the embedding manifold. In 3D there are additional local fields describing the director, i.e. fields available to a local observer residing within the material, and a number of distinct ways to yield geometric frustration. So far it was unknown how many such local fields are required to uniquely describe a 3D director field, nor what are the compatibility relations they must satisfy. In this work, we address these questions directly. We employ the method of moving frames to show that a director field is fully determined by five local fields. These fields are shown to be related to each other and to the curvature of the embedding space through six differential relations. As an application  of  our  method,  we characterize all uniform  distortion director  fields,  i.e.,  directors  for  which  all the local characterizing fields are  constant in space, in manifolds of constant curvature. The  classification  of  such  phases  has  been recently provided for directors in Euclidean space, where the textures correspond to foliations of space by parallel congruent helices. For non-vanishing curvature, we show that the pure twist phase is the only solution in {positively curved space}, while in the hyperbolic space uniform distortion fields correspond to foliations of space by (non-necessarily parallel) congruent helices. Further analysis of the obtained compatibility fields is expected to allow to also construct new non-uniform director fields.
\end{abstract}

%

%
%
%
%

\section{Introduction}
\label{intro}
Liquid crystals are a state of matter characterized by the presence of an orientational order but no, or only partial, positional order. In many cases, the ordering can be described in terms of a unit vector field $\mathbf{n}$, called the director \cite{GP95,Virga1995}. Liquid crystals pervade our daily lives, from computer and smart-phone displays to optical switches enabling fast and efficient communication. In recent years, liquid crystals also found applications as controllable and responsive materials \cite{ASK14,GAE19,WM18,SMNML16}, and similar phases were identified outside to soft matter systems, for example in the nematic order observed in Iron based superconductors \cite{FCS14,WKL15}.

The liquid crystalline orientational ``texture" often manifests the shape and interactions between its constituents. Elongated and straight constituents with steric interactions favor the nematic phase in which the director's orientation is uniform in space. In contrast, chiral constituents may favor a twisted director field, while elongated and curved constituents may favor a bent director. However, not all such locally preferred tendencies can be globally realized by a director field in a finite domain. For example, the two dimensional straight nematic texture with vanishing splay and bend cannot be realized on any open region on {the surface of a sphere} \cite{NivSM2018}. Here, the splay and bend of a director field $\mathbf{n}$ are given by $s=\nabla\cdot\mathbf{n}$ and $b=\Vert(\mathbf{n}\cdot\nabla)\mathbf{n}\Vert$, respectively, and constitute the basic distortion modes of any two dimensional director field. Similarly, the phase of constant non-vanishing bend and vanishing splay cannot be realized in the plane \cite{Meyer1976}. It is thus natural to ask what local tendencies could be realized by a director texture, and conversely how many such local descriptors are required to uniquely determine a texture. 

Recently, it was shown that any two dimensional director field is fully described by its bend and splay fields, and that the values these scalar fields obtain for any realizable texture satisfy
$K = -b^2-s^2-\mathbf{n}\cdot\nabla s+\mathbf{n}_{\perp}\cdot\nabla b$ \cite{NivSM2018}, where $K$ is the Gaussian curvature of the surface $S$ in which the field is embedded and $\mathbf{n}_{\perp}$ is the field in $S$ normal to $\mathbf{n}$. The identification of the class of all admissible textures also allowed addressing the notion of optimal compromise for unrealizable frustrated states. These results are, however, presently limited to two dimensional systems.
For three dimensional liquid crystals there are additional distortion fields, such as the twist and saddle-splay that do not have corresponding fields in two dimensional systems. Moreover, the three dimensional geometry is associated with additional compatibility conditions; while for two dimensional Riemannian geometry there is only one local geometric charge, in three dimensions there are three independent scalar Riemannian charges. Thus, the three dimensional case is expected to lead to a larger set of relations involving a greater number of fields. Presently, it is unknown how many fields are required to uniquely determine a director field in three dimensions, nor how many relations these fields must satisfy to correspond to a realizable texture. 

Many frustrated assemblies, in which the constituents locally favor an arrangement that cannot be globally realized, exhibit a super-extensive ground state energy for isotropic domains; i.e. the energetic cost of the optimal compromise in these systems increases faster than linearly with their mass. Recently, in was shown that the exact order and structure of the compatibility conditions completely determines this super-extensive behavior and can be used to predict the exponent related to the super-extensive growth of the ground state energy \cite{ME21}. 
The purpose of the work presented here is to further advance recent efforts aimed at understanding and quantifying frustration in three dimensional liquid crystals. We provide a definitive answer to the above questions by writing explicitly the six differential relations that form the compatibility conditions relating the five fields that describe a director field in three dimensions. These six equations relate the fields and their derivatives to each other and to the curvature tensor of the 3D manifold where the director field lives. As an application of our results, we also characterize all uniform distortion fields in the {three-sphere, $\mathbb{S}^3$}, and hyperbolic space, {$\mathbb{H}^3$}, showing in particular that in hyperbolic space uniform distortion fields also correspond to a foliation of space by (non-necessarily parallel) helices\footnote{By a helix  we mean a curve  of constant curvature and constant torsion.}. Thus, together with the results of reference \cite{VirgaPRE2019}, we complete the characterization of uniform distortion fields for all the three homogeneous and isotropic geometries. 

{For flat space, satisfaction of the compatibility conditions constitutes a necessary and sufficient condition for the existence of a corresponding director field. We thus conclude that knowledge of the five scalar fields that describe the director; namely the twist, $t$, splay $s$, bend $b$, biaxial splay $\Delta$, and relative orientation between the principal biaxial splay direction and the bend direction $\phi$, suffice for defining a texture, unique up to rigid motions, provided that they satisfy the compatibility conditions.}

\section{Background: Geometric frustration in three-dimensional director fields}
\label{intro2}

The present work joins ongoing efforts to better understand the underlying geometry of three dimensional director fields. Recent insightful interpretations of the basic distortion modes of unit director fields in three dimensions identified these distortion modes with distinct components of the director gradient, $J=\nabla\mathbf{n}$ \cite{MachonPRX2016,SelingerLCR2018}. The splay corresponds to the trace of $J$, while the bend is a vector in the space perpendicular to $\mathbf{n}$ and thus contributes two degrees of freedom. The remaining modes contribute to the components of $J$ in the two dimensional space normal to $\mathbf{n}$, and are traceless.
The twist $t=\mathbf{n}\cdot(\nabla\times\mathbf{n})$ corresponds to the anti-symmetric component, while the biaxial splay is identified with the remaining traceless symmetric structure and thus contributes two degrees of freedom as well. This yields a total count of six independent contributions to $J$ \cite{MachonPRX2016,SelingerLCR2018}. However, the freedom in assigning a base to the space perpendicular to $\mathbf{n}$ eliminates one of these to yield five total intrinsic fields that describe a director. We identify these as the splay, bend, twist, saddle-splay and the relative orientation between the direction of the bend vector and the principal direction of the biaxial splay.

These local descriptors of the liquid crystalline order may be associated with non trivial reference values induced by the structure and relative interactions of their constituents. Considering phases composed of identical constituents, it is natural to assume that these reference values will be uniform in space and manifest the underling symmetry of their constituent. However, as was recently shown \cite{VirgaPRE2019}, the space of phases associated with such constant descriptors, termed ``uniform distortions", is very limited, necessitating more complex textures. For example, chiral constituents favoring the unrealizable uniform double twist produce the Blue phase in which defect lines, separating biaxially twisted columns, are periodically arranged \cite{GP95}. Similarly, achiral bent core liquid crystals form chiral meso-phases displaying giant optical activity \cite{B+13,H+09} and heliconical ordering \cite{C+14}.  The constituents in this case locally favor a phase of vanishing twist, splay and saddle-splay and a constant non-vanishing bend. Such a phase cannot be realized in Euclidean space and instead the system incorporates a twist in order to accommodate the uniform bend resulting in the observed heliconical phase \cite{Meyer1976}. 

Focusing on uniform distortions Virga showed that all such textures correspond to foliations of the three dimensional Euclidean space by parallel helices \cite{VirgaPRE2019}. His results relied on vector calculus, where the motion of the frame $\{\mathbf{n}=\mathbf{n}_1,\mathbf{n}_2,\mathbf{n}_3\}$ is described in terms of the so-called connectors vector fields and the compatibility conditions for the deformation modes associated with a director then follow from the symmetry of the tensors $\mathbf{n}_i\cdot\nabla^2\mathbf{n}_j$ \cite{VirgaPRE2019}. In particular, it was shown that the pure bend phase favored by bent core liquid crystals is indeed frustrated, and predicted the heliconical phase with uniform twist as a plausible compromise. For small enough domains, however, one might expect other non-uniform distortions to yield the optimal compromise \cite{ME21}. 

Similar arguments show that the attempted pure double twist phase resulting in the blue phase is also frustrated in Euclidean space \cite{VirgaPRE2019}. This attempted phase, however, can be accommodated in a three-dimensional spherical geometry of an appropriate radius \cite{SethnaPRL1983}. Other examples of uniform distortion fields have been recently provided for all the eight Thurston geometries \cite{SadocNJP2020}, where it is shown that each pure mode of director deformation can fill space without frustration for at least one type of geometry.

In this work we seek to obtain the full compatibility conditions for three dimensional director fields. Naturally, one may seek to exploit the same reasoning that was exploited to yield the compatibility conditions in two dimensions \cite{NivSM2018}. However, the method employed there relies heavily on the existence of a natural orthogonal frame of coordinates such that the parametric curves are tangent to $\mathbf{n}$ and to the perpendicular unit vector $\mathbf{n}_{\perp}$. This, however, could not be generalized to three dimensions. A general field of an orthonormal triad in 3D cannot be associated with the tangents of parametric curves. Instead, one needs to study the properties of the orthonormal triad field without resorting to coordinates; the mathematical formalism which achieves this is called {\em the method of moving frames} \cite{Clelland2017}, also known as vielbein formalism in the context of relativity \cite{Car03}. Given a 3D director field $\mathbf{n}_1=\mathbf{n}$ and its two normals $\mathbf{n}_{2}$ and $\mathbf{n}_3=\mathbf{n}_1\times\mathbf{n}_{2}$ one can build the corresponding dual frame  of differential forms which together with the so-called connection forms describe the geometry of 3D space using the differential form structure equations. This formalism also allows for an invariant formulation of vector calculus operators, which means that quantities and energy functionals used in the description of 3D liquid crystals can be rewritten as exterior differential systems, i.e., differential equations in terms of differential forms and operations defined on them.

Though more abstract than the vector calculus method \cite{VirgaPRE2019}, the approach based on differential forms allows us to obtain manageable equations and to investigate director fields in both Euclidean and curved Riemannian spaces in an equal foot. This helps in better understanding how the Euclidean space frustrates the existence of certain phases. 

When concluding the writing of this manuscript a parallel effort to obtain the compatibility conditions using moving frames by Pollard and Alexander came to our attention \cite{PollardArXiv2021}.  We briefly relate to the similarities and differences between these works in the discussion section.

\section{Differential forms and moving frames}

Given a coordinate system $(x^1,\dots,x^m)$ on an open and connected set $U\subseteq\mathbb{R}^m$, the corresponding vector fields tangent to the coordinate curves are denoted by $\{\frac{\partial}{\partial x^i}\}$, while their dual fields (or \emph{covectors}) are denoted by $\rmd x^i$, i.e., when applied to a vector $v=v^i\frac{\partial}{\partial x^i}$ (sum on repeated indices), we have $\rmd x^i(v)=v^i$. 

The differential of a scalar function $f$ is defined as $\rmd f=\frac{\partial f}{\partial x^i}\rmd x^i$ and, consequently, $\rmd x^i$ can be alternatively seem as the differential of the $i$-th coordinate function. From now on, a field of covectors $p\in U\mapsto \eta_p\in(T_pU)^*$ is called a \emph{differential 1-form}, while a function is a \emph{0-form}. Notice that we can write any 1-form as $\eta=a_i\rmd x^i$ for some scalar fields $a_i$ and that there is an isomorphism between vector fields and 1-forms: $a_i\rmd x^i\leftrightarrow a_i\frac{\partial}{\partial x^i}$. Given two differential 1-forms $\eta$ and $\omega$, we define the \emph{exterior product} $\eta\wedge\omega$ as the anti-symmetric bilinear map $(\eta\wedge\omega)(u,v)=\eta(u)\omega(v)-\eta(v)\omega(u)$. We shall refer to $\eta\wedge\omega$ as a \emph{differential 2-form}. We can define the \emph{exterior derivative} of a 1-form $\eta=a_i\rmd x^i$ as the 2-form $\rmd\eta=\rmd a_i\wedge\rmd x^i=\frac{\partial a_i}{\partial x^j}\rmd x^j\wedge\rmd x^i$. The vector space of 2-forms are generated by $\{\rmd x^i\wedge\rmd x^j\}_{1\leq i<j\leq m}$ and, therefore, it has dimension $m(m-1)/2$. More generally, a \emph{differential $k$-form} is an anti-symmetric $k$-linear map and the corresponding vector space is generated by the basis $\{\rmd x^{i_1}\wedge\dots\wedge\rmd x^{i_k}\}_{1\leq i_1<\dots<i_k\leq m}$, where $(\rmd x^{i_1}\wedge\dots\wedge\rmd x^{i_k})(v_1,\dots,v_m)=\det(\rmd x^{i_r}(v_s))_{rs}$. The \emph{exterior derivative} of a $k$-form $\eta=a_{i_1\dots i_k}\rmd x^{i_1}\wedge\dots\rmd x^{i_k}$ is the $(k+1)$-form $\rmd\eta=\rmd a_{i_1\dots i_k}\wedge\rmd x^{i_1}\wedge\dots\rmd x^{i_k}$. In addition, $\rmd$ is linear and satisfies the product rule $\rmd(\eta\wedge\omega)=\rmd\eta\wedge\omega+(-1)^{k}\eta\wedge\rmd\omega$, where $\eta$ is a $k$-form and $\omega$ is a $\ell$-form. A remarkable property of the exterior derivative $\rmd$ is that $\rmd^2=0$, i.e., the differential of the $k$-form $\rmd\eta$ always vanishes. (As an exercise, the reader can easily verify this property for 0- and 1-forms.) 

Instead of using coordinate fields, we may consider in $U$ any set of orthonormal vector fields $\{\mathbf{n}_1,...,\mathbf{n}_m\}$ along with its set of dual fields $\{\eta^1,...,\eta^m\}$, i.e., $\eta^i(\mathbf{n}_j)=\delta_j^i$, where $\delta_j^i$ is the Kronecker delta. Since each $\mathbf{n}_i$ is a smooth map from $U$ to $\mathbb{R}^m$, if we write it in coordinates $\mathbf{n}_i=n_i^j\frac{\partial}{\partial x^j}$, its differential\footnote{The use of the same symbol for both the differential of a map between manifolds and the exterior derivative of a differential form is justified by the possibility of seeing the differential as a vector-valued 1-form, see, e.g. subsection 2.8 of reference \cite{Clelland2017}.} is $\rmd\mathbf{n}_i=(\rmd n_i^1,\dots,\rmd n_i^m)=\frac{\partial\mathbf{n}_i}{\partial \mathbf{x}}\rmd \mathbf{x}$, where $\frac{\partial\mathbf{n}_i}{\partial \mathbf{x}}$ is the Jacobian matrix acting by matrix multiplication on $\rmd\mathbf{x}=(\rmd x^1,\dots,\rmd x^m)$. Alternatively, the differential $\rmd\mathbf{n}_i$ acting on a tangent vector $v\in T_pU$ can be written as a linear combination
\begin{equation}
    (\rmd\mathbf{n}_i)_p(v) = \eta_i^j(p,v)\,\mathbf{n}_j(p).
\end{equation}
In what follows, we shall omit the explicit dependence on $p$ and $v$ and simply write $\rmd\mathbf{n}_i= \eta_i^j\mathbf{n}_j$. For a fixed point $p$, the functions $\eta_i^j$ are linear and, therefore, each $p\mapsto \eta_i^j(p,\cdot)$ defines a 1-form. From the orthonormality of $\{\mathbf{n}_i\}$ it follows that $\eta_i^j=-\eta_j^i$.  

If $\mathbf{r}:U\to\mathbb{R}^m$ denotes the inclusion map, its differential can be written as $\mathrm{d}\mathbf{r} = \eta^i\,\mathbf{n}_i$. Geometrically, given a moving frame $\{\mathbf{n}_i\}$, the set of 1-forms $\{\eta^i\}$ describes infinitesimal translations of the moving frame while the 1-forms $\{\eta_i^j\}$ describes infinitesimal rotations. Now, using that $\rmd^2\mathbf{r}=0$ and $\rmd^2\mathbf{n}_i=0$, we have the so-called \emph{structure equations}
\begin{equation}
\left\{
\begin{array}{c}
\mathrm{d}\eta^i = \eta^k\wedge\eta_k^i\\[5pt]
\mathrm{d}\eta_j^i = \eta_j^k\wedge\eta_k^i\\
\end{array}
\right.,\,i,j\in\{1,\dots,m\}.\label{eq::StructureEqs}
\end{equation}
For the Euclidean case, these constitute the integrability conditions
for the existence of a moving frame with dual frame $\{\eta^i\}$ and connection forms $\{\eta_i^j\}$ \cite{G74}. See \cite{T71}, lemma 2 with $k=0$, for an elementary proof.  

As an example of these ideas, consider in $\mathbb{R}^3$ the moving frame given by the vector fields $\mathbf{n}_1=(\cos\phi\cos\theta,\cos\phi\sin\theta,\sin\phi)$, $\mathbf{n}_2=(-\sin\theta,\cos\theta,0)$ and $\mathbf{n}_3=(-\sin\phi\cos\theta,-\sin\phi\sin\theta,\cos\phi)$, where $\theta=\theta(x,y,z)$ and $\phi=\phi(x,y,z)$ are smooth functions on $U\subseteq\mathbb{R}^3$. Computing their differential gives
\begin{eqnarray*}
     \rmd\mathbf{n}_1 & = & \rmd\phi(-\sin\phi\cos\theta,-\sin\phi\sin\theta,\cos\phi)+\rmd\theta(-\cos\phi\sin\theta,\cos\phi\cos\theta,0)\nonumber\\
     & = & \cos\phi\,\rmd\theta\,\mathbf{n}_2+\rmd\phi\,\mathbf{n}_3,\\
     \rmd\mathbf{n}_2 & = & \rmd\theta(-\cos\theta,-\sin\theta,0)= -\cos\phi\,\rmd\theta\,\mathbf{n}_1+\sin\phi\,\rmd\theta\,\mathbf{n}_3,\\
     \rmd\mathbf{n}_3 & = & -\rmd\phi(\cos\phi\cos\theta,\cos\phi\sin\theta,\sin\phi)+\rmd\theta(\sin\phi\sin\theta,-\sin\phi\cos\theta,0)\nonumber\\
     & = &-\rmd\phi\,\mathbf{n}_1- \sin\phi\rmd\theta\,\mathbf{n}_2.
\end{eqnarray*}
Therefore, the 1-forms $\eta_i^j$ associated with $\{\mathbf{n}_1,\mathbf{n}_2,\mathbf{n}_3\}$ are $\eta_1^2=\cos\phi\,\rmd\theta$, $\eta_1^3=\rmd\phi$, and $\eta_2^3=\sin\phi\,\rmd\theta$. We leave as an exercise checking the validity of the structure equations $\rmd\eta_j^i=\eta_j^k\wedge\eta_k^i$.

The 1-forms $\eta_i^k$ are also known as \emph{connection forms} since they determine the connection coefficients of the covariant derivative. Indeed, given two vector fields $u=u^i\mathbf{n}_i$ and $v=v^i\mathbf{n}_i$ in $U$, the covariant derivative of $u$ in the direction of $v$, $\nabla_vu$, can be written using moving frames as
\begin{equation}
\nabla_vu = (\rmd u)(v)=\rmd(u^k\mathbf{n}_k)(v)=\left[\mathrm{d}u^k(v)+u^j\eta_j^k(v)\right]\,\mathbf{n}_k\,.
\end{equation}
Therefore, the connection forms can be alternatively computed from the Levi-Civita connection $\nabla$ by using the relation $\eta_j^k(v)=\langle\nabla_{v}\mathbf{n}_j,\mathbf{n}_k\rangle$. In addition, given two tangent vectors $u,v\in T_pU$, the inner product between them is $g(u^i\mathbf{n}_i,v^j\mathbf{n}_j)=u^iv^j\delta_{ij}=u^iv^i=\eta^i(u)\eta^i(v)$. The metric $g$ in $U$ is then written as $g=(\eta^1)^2+\dots+(\eta^m)^2$. It follows that  the geometry of $U\subseteq\mathbb{R}^m$ is entirely contained in the sets of 1-forms $\{\eta^i\}$ and $\{\eta_i^j\}$.

To accomplish the goal of doing differential geometry using moving frames, we should be able to compute differential operators using differential forms. To do that, we need the Hodge star operator $\star$, which takes $k$-forms to $(m-k)$-forms. Geometrically, we proceed as follows. Given a $k$-form $\omega=\omega^1\wedge\dots\wedge\omega^k$, where $\{\omega^i\}$ is linearly independent, consider the $k$-dimensional vector subspace $V$ of $\mathbb{R}^m$ generated by the vectors $\{v_{1},\dots,v_k\}$ associated with $\{\omega^1,\dots,\omega^k\}$. We then pick a basis $\{v_{k+1},\dots,v_m\}$ of the vector space $V^{\perp}$ orthogonal to $V$ and consider $\omega^{k+1},\dots,\omega^m$, the 1-forms associated with the vectors of this basis. Then, we define $\star\,\omega=\pm\lambda\,\omega^{k+1}\wedge\dots\wedge\omega^m$, where $\lambda$ is the $k$-volume of the solid generated by $\{v_i\}_{i=1}^k$ and the sign corresponds to the orientation of $\mathcal{B}=\{v_1,\dots,v_k,v_{k+1},\dots,v_m\}$, i.e., plus if $\mathcal{B}$ has the same orientation as the canonical basis of $\mathbb{R}^m$ and minus if otherwise. Finally, we compute $\star$ for a generic linear form by demanding linearity. As an example, in $\mathbb{R}^3$ the Hodge star operator acting on 1-forms gives $\star\rmd x^1=\rmd x^2\wedge \rmd x^3$, $\star\rmd x^2=-\rmd x^1\wedge \rmd x^3$, and $\star\rmd x^3=\rmd x^1\wedge \rmd x^2$. Finally, the curl and divergence of $\mathbf{n}$ are associated with differential forms according to
\begin{equation}
\nabla\times\hat{\mathbf{n}} \leftrightarrow \star(\mathrm{d}\eta)\mbox{ and }
\nabla\cdot\hat{\mathbf{n}} = \star[\mathrm{d}(\star\,\eta)],
\end{equation}
where $\eta$ is the 1-form dual to $\hat{\mathbf{n}}$.

\section{Compatibility condition for two-dimensional director fields}

Director fields $\mathbf{n}$ in 2D are fully described by their bend $b=\Vert\mathbf{n}\times\nabla\times\mathbf{n}\Vert$ and splay $s=\nabla\cdot\mathbf{n}$. However, the splay and bend are not independent functions and they are related to the curvature of the ambient surface by \cite{NivSM2018} $$-K=s^2+b^2+\mathbf{n}\cdot\nabla s-\mathbf{n}^{\perp}\cdot\nabla b.$$ 
In this section we provide an alternative proof for the 2D compatibility equation via moving frames. But, first, we shall illustrate how the moving frame method can be used to describe the geometry of surfaces.

Let $\mathbf{r}:U\to S\subset\mathbb{R}^3$ be a surface and $\mathbf{N}$ its unit normal. If $\{\mathbf{n}_1,\mathbf{n}_2\}$ is a field of orthonormal bases for the tangent planes, we then define a moving frame  in 3D as $\{\mathbf{n}_1,\mathbf{n}_2,\mathbf{n}_3:=\mathbf{N}\}$ along with its dual frame $\{\eta^1,\eta^2,\eta^3\}$. Since we are interested on the surface geometry, we shall restrict our attention to $\eta^i,\eta_i^j$ when applied to tangent vectors. Then, in this restricted setting it follows that
\[
\forall\,v=v^1\mathbf{n}_1+v^2\mathbf{n}_2\in T_pS,\,\eta^3(v)=0.
\]
Therefore, seeing $\eta^3$ as a 2D differential form on $S$ implies $\eta^3=0$. Thus, the 1-forms $\eta_1^2,\eta_1^3,\eta_2^3$ can be written as a linear combination of $\eta^1$ and $\eta^2$ only, i.e., they can also be seen as differential forms on the surface. This process of seeing $\eta^i$ and $\eta_j^i$ as 2D differential forms can be rigorously justified by using $\mathbf{r}$ to pullback the 1-forms $\eta^i$ and $\eta_i^j$ to $U$: the pullback of a $k$-form $\eta$ is the $k$-form $\omega=\mathbf{r}^*\eta$ defined by $\omega_p(v_1,\dots,v_k)=\eta_{\mathbf{r}(p)}(\rmd\mathbf{r}(v_1),\dots,\rmd\mathbf{r}(v_k))$. Now, since the pullback operation $^*$ commutes with $\rmd$ and $\wedge$ \cite{doCarmo1994DiffForms}, the 1-forms $\mathbf{r}^*\eta^i$ and $\mathbf{r}^*\eta_j^i$ satisfy the same structure equations as $\eta^i$ and $\eta_j^i$. Thus, with some abuse of notation, we simply write $\eta^i=\mathbf{r}^*\eta^i$ and $\eta_j^i=\mathbf{r}^*\eta_j^i$, which finally justifies seeing $\eta^i$ and $\eta_i^j$ as 1-forms over $S=\mathbf{r}(U)$ (\footnote{As an alternative to using pullbacks, we could consider a foliation of space by surfaces parallel to $S$ spanning a region parametrized as $\mathbf{R}(x^1,x^2,x^3)=\mathbf{r}(x^1,x^2)+x^3\mathbf{N}(x^1,x^2)$. Since we are only interested on tangent directions, any dependence of $\eta_i^j$ on $\eta^3$ does not contribute to the final result. In addition, following this idea, $\eta^3$ is nothing but the differential of the $x^3$-coordinate, which implies that $\rmd\eta^3=0$. As shown in the main text, this is the key property allowing us to use the moving frame method to study the differential geometry of surfaces in space.}).   

From the fact that $\eta^3=0$ on $S$, it follows that $\mathrm{d}\eta^3=0$ on $S$. Then, the structure equations in (\ref{eq::StructureEqs}) imply that $\eta^1\wedge\eta_1^3+\eta^2\wedge\eta_2^3=0.$ An important result for differential forms is the Cartan lemma \cite{doCarmo2012MovingFrames,Clelland2017}, which says that if $\omega^1,...\,,\omega^k$ are linearly independent 1-forms and if there exist 1-forms $\theta^1,...\,,\theta^k$ such that $\sum_{i=1}^k\omega^i\wedge\theta^i=0$, then $\theta^i = a_j^i\,\omega^j$ with $a_i^j=a_j^i$. Therefore, since the set $\{\eta^1,\eta^2\}$ is linearly independent, from the Cartan lemma we may write
\begin{equation}\label{eq::SurfWeingartenEqsViaMovingFrames}
\eta_1^3 = a_1^1\eta^1+a_2^1\eta^2\mbox{ and }
\eta_2^3 = a_1^2\eta^1+a_2^2\eta^2,\,a_i^j=a_j^i.
\end{equation}
From $\mathrm{d}\mathbf{n}_3=\eta_3^1\,\mathbf{n}_1+\eta_3^2\,\mathbf{n}_2=-(\eta^3_1\,\mathbf{n}_1+\eta^3_2\,\mathbf{n}_2)$, it follows that the coefficients $a_j^i$ precisely describe the shape operator of $S$. Then, the mean ($H$) and Gaussian ($K$) curvatures  can be written as 
\begin{equation}\label{eq::KandMviaMovingFrames}
H = \frac{1}{2}\mathrm{tr}(a)=\frac{a_1^1+a_2^2}{2}\mbox{ and }K = \det(a)=a_1^1a_2^2-(a_1^2)^2.
\end{equation}

It remains to find the interpretation of $\eta_1^2$. From $\eta_1^2(v)=\langle \nabla_v\mathbf{n}_1,\mathbf{n}_2\rangle$, we see that we can write $\eta_1^2=\eta_1^2(\mathbf{n}_1)\eta^1+\eta_1^2(\mathbf{n}_1)\eta^2=\kappa_g\eta^1+\kappa_g^{\perp}\eta^2$, where $\kappa_g$ and $\kappa_g^{\perp}$ are the geodesic curvatures of the integral curves of $\mathbf{n}_1$ and $\mathbf{n}_2$, respectively. In addition, taking the exterior derivative provides the important relation $\rmd \eta_1^2=\eta_1^3\wedge\eta_1^2=-K\eta^1\wedge\eta^2$. This relation will be the key to finding the compatibility equation for director fields in 2D.

We have just seen that for a surface in 3D the intrinsic geometry is encoded in $\eta^1,\eta^2$, and $\eta_1^2$, while the extrinsic geometry comes from $\eta_1^3$ and $\eta_2^3$. (The second fundamental form $\mathrm{II}$ can be written as $\mathrm{II}=\eta^i\eta_i^3$.) The equation $\rmd \eta_1^2=-K\eta^1\wedge\eta^2$ in 2D indicates that for moving frames in a Riemannian manifold the second set of structure equations, equation \eref{eq::StructureEqs}, must be modified to account for the curvature of the ambient manifold: For a 2D manifold with Gaussian curvature $K=R_{1212}$, the structure equations associated with the 1-forms $\{\eta^1,\eta^2\}$ and $\eta_1^2$ are $\rmd\eta^1=\eta^2\wedge\eta_2^1$, $\rmd\eta^2=\eta^1\wedge\eta_1^2$, and $\rmd\eta_1^2-\eta_1^k\wedge\eta_k^2=\rmd\eta_1^2=-K\eta^1\wedge\eta^2$.

In general, for a moving frame $\{\mathbf{n}_i\}_{i=1}^m$ in a Riemannian manifold $M^m$ with curvature tensor $R_{ijk\ell}=R_{jk\ell}^i$, the structure equations are \cite{doCarmo2012MovingFrames,Clelland2017}
\begin{equation}\label{eq::CurvedStructureEqs}
\rmd\eta^i=\eta^k\wedge\eta_k^i \mbox{ and }\rmd\eta_j^i-\eta_j^k\wedge\eta_k^i=-\frac{1}{2}R_{jk\ell}^i\eta^{k}\wedge\eta^{\ell}=-\sum_{k<\ell}R_{ijk\ell}\eta^{k}\wedge\eta^{\ell},
\end{equation}
where we used that the operation of raising and lowering indices is trivial since the metric coefficients associated with the moving frame are $\delta_{ij}=\langle\mathbf{n}_i,\mathbf{n}_j\rangle$. 

Now, let $\hat{\mathbf{n}}$ be a director field on a 2D Riemannian manifold $(M^2,\langle\cdot,\cdot\rangle)$. We may introduce a moving frame $\{\mathbf{n}_1:=\hat{\mathbf{n}},\mathbf{n}_2:=\hat{\mathbf{n}}^{\perp}\}$ along with its coframe $\{\eta^1,\eta^2\}$. As we have seen, we can write $\eta_1^2 = \kappa_g\,\eta^1+\kappa_g^{\perp}\,\eta^2$.

On the one hand, the splay $s=\nabla\cdot\hat{\mathbf{n}}$ is computed as
\begin{equation}
s = \star\,\mathrm{d}\star\eta^1=\star\,\mathrm{d}\eta^2=\star\,(\eta^1\wedge\eta_1^2) = \kappa_g^{\perp}.
\end{equation}
On the other hand, the bend $b=\Vert\hat{\mathbf{n}}\times\nabla\times\hat{\mathbf{n}}\Vert$ is
\begin{eqnarray}
b & = & \left\Vert\star(\eta^1\wedge\star\,\mathrm{d}\eta^1)\right\Vert=\left\Vert\star\,[\,\eta^1\wedge\star\,(\eta^2\wedge\eta_2^1)]\,\right\Vert\nonumber\\
& = & \left\Vert\star\,[\,\eta^1\wedge\star\,(\kappa_g\,\eta^1\wedge\eta^2)]\,\right\Vert=\left\Vert\kappa_g\eta^2\right\Vert=\kappa_g.
\end{eqnarray}
This last equation also shows that, in 2D, we may write $b=-\nabla\cdot\mathbf{n}_2=\star\,\mathrm{d}\star\eta^2$. In short, we have the following relation
\begin{equation}
\eta_1^2=b\,\eta^1+s\,\eta^2.\label{eq::Omega12InTermsOfsplayAndBend}
\end{equation}

Now we shall apply the findings above in order to write the compatibility equation for 2D director fields as found in \cite{NivSM2018}, but using moving frames.

\begin{theorem}[Compatibility condition in 2D]
Let $\hat{\mathbf{n}}$ be a director field  with splay $s$ and bend $b$ on a 2D manifold $M^2$ with Gaussian curvature $K$. Then, 
\begin{equation}
-K= s^2+b^2+(\hat{\mathbf{n}}\cdot\nabla)\,s-(\hat{\mathbf{n}}^{\perp}\cdot\nabla)\,b,
\end{equation}
where $(v\cdot\nabla)$ is the directional derivative in the direction of $v$ and $\langle\hat{\mathbf{n}},\hat{\mathbf{n}}^{\perp}\rangle=0$.
\end{theorem}
\textit{Proof.} The exterior derivative of $\eta_1^2$ is
\begin{eqnarray*}
\mathrm{d}\eta_1^2 & = & \mathrm{d}(b\,\eta^1+s\,\eta^2)= \mathrm{d}b\wedge\eta^1+b\,\mathrm{d}\eta^1+\mathrm{d}s\wedge\eta^2+s\,\mathrm{d}\eta^2\nonumber\\
& = & [(\hat{\mathbf{n}}^{\perp}\cdot\nabla)b]\,\eta^2\wedge\eta^1+b\,\eta^2\wedge\eta_2^1+[(\hat{\mathbf{n}}\cdot\nabla)s]\,\eta^1\wedge\eta^2+s\,\eta^1\wedge\eta_1^2\nonumber\\
& = & \left[(\hat{\mathbf{n}}\cdot\nabla)\,s-(\hat{\mathbf{n}}^{\perp}\cdot\nabla)\,b+b^2+s^2\right]\,\eta^1\wedge\eta^2.
\end{eqnarray*}
Now, using that $\mathrm{d}\eta_1^2=-K\,\eta^1\wedge\eta^2$ we deduce the desired equality.

\section{Three-dimensional director fields}

Inspired by the study of 2D director fields, the strategy in 3D will consist of writing the 1-forms $\eta_i^j$ in terms of the deformation modes of a director field $\mathbf{n}$ and then from the structure equations associated with $\rmd\eta_i^j$ we will obtain the compatibility equations. 

In 2D, there are two deformation modes (bend and splay), while in 3D there are 6 modes, which can be further reduced to 5. Indeed, as discussed in Sect. \ref{intro2}, taking into account rotations that preserve the director $\mathbf{n}$, the gradient $\nabla\mathbf{n}$ decomposes as \cite{MachonPRX2016,SelingerLCR2018}
\begin{equation}
    \nabla_{\alpha}n_{\beta} = -n_{\alpha}\,b_{\beta}+\frac{s}{2}(\delta_{\alpha\beta}-n_{\alpha}n_{\beta})+\frac{t}{2}\epsilon_{\alpha\beta\gamma}n_{\gamma}+\Delta_{\alpha\beta},
\end{equation}
where Greek indices indicate Cartesian coordinates and $\mathbf{b}=-\mathbf{n}\cdot\nabla\mathbf{n}$ is the \emph{bend vector}, whose norm $b$ (the \emph{bend}) gives the curvature of the integral lines of $\mathbf{n}$, $s=\nabla\cdot\mathbf{n}$ is the \emph{splay}, $t=\mathbf{n}\cdot\nabla\times\mathbf{n}$ is the \emph{twist}, and $\Delta_{ij}$ are the coefficients of the \emph{biaxial-splay} \cite{SelingerLCR2018}.

In 2D, the coefficients of the 1-form $\eta_1^2$ are related to the geometry of the integral curves of the director and its orthogonal field. Given an integral curve of $\mathbf{n}_i$ in 3D, we can consider $\{\mathbf{n}_i,\mathbf{n}_{i+1},\mathbf{n}_{i+2}\}$ as a positive orthonormal moving trihedron along it, e.g., for $\mathbf{n}_2$ we have $\{\mathbf{n}_2,\mathbf{n}_3,\mathbf{n}_1\}$. The equations of motion of such a moving trihedron along the $\mathbf{n}_i$-integral curves are
\begin{equation}\label{eq::EqsOfMotionMovFrame}
\nabla_{\mathbf{n}_i}\left(
\begin{array}{c}
\mathbf{n}_i\\
\mathbf{n}_{i+1}\\
\mathbf{n}_{i+2}\\
\end{array}\right)=\left(
\begin{array}{ccc}
0 & \kappa_i^1 & \kappa_i^2\\[4pt]
-\kappa_i^1 & 0 & \omega_i\\[4pt]
-\kappa_i^2 & -\omega_i & 0\\
\end{array}\right)\left(
\begin{array}{c}
\mathbf{n}_i\\
\mathbf{n}_{i+1}\\
\mathbf{n}_{i+2}\\
\end{array}\right),
\end{equation}
where $\kappa_i^1$ and $\kappa_i^2$ relate to the (geodesic) curvature function $\kappa_i$ as $\kappa_i(s)=\sqrt{[\kappa_i^1(s)]^2+[\kappa_i^2(s)]^2}$ and $\omega_i$ relates to the torsion $\tau_i$ as $\omega_i(s)=\tau_i(s)-\theta'(s)$, where $\theta$ is the angle between the (Frenet) principal normal and $\mathbf{n}_{i+1}$ \cite{TakagiPTP1992}. Thus, using the property $\eta_i^j(v)=\langle\nabla_{v}\mathbf{n}_i,\mathbf{n}_j\rangle$, the 1-forms $\eta_i^j$ when written in the basis $\{\eta^1,\eta^2,\eta^3\}$ are
\begin{equation}
\left\{
\begin{array}{c}
\eta_1^2 = \kappa_1^1\,\eta^1-\kappa_2^2\,\eta^2+\omega_3\,\eta^3\\[4pt]
\eta_1^3 = \kappa_1^2\,\eta^1-\omega_2\,\eta^2-\kappa_3^1\,\eta^3\\[4pt]
\eta_2^3 = \omega_1\,\eta^1+\kappa_2^1\,\eta^2-\kappa_3^2\,\eta^3\\
\end{array}
\right..\label{eq::omegaijInTermsOfKappaijAndWi}
\end{equation}

The 1-forms $\eta_1^2$ and $\eta_1^3$ provide information about the gradient of the director $\mathbf{n}_1$, $\rmd\mathbf{n}_1=\eta_1^2\mathbf{n}_2+\eta_1^3\mathbf{n}_3$. The components dual to the director $\mathbf{\hat{n}}$ then contains information about the bend vector $\mathbf{b}=-\nabla_{\mathbf{n}}{\mathbf{n}}=b\,\mathbf{n}\times\nabla\times\mathbf{n}$, $b = \sqrt{(\kappa_1^1)^2+(\kappa_1^2)^2}$. The remaining components of $\rmd\mathbf{n}_1$ can be decomposed into an antisymmetric and a symmetric part, where the symmetric part can be further decomposed into a trace and traceless operator. This decomposition provides the twist $t$, splay $s$, and biaxial splay coefficients $\Delta_{ij}$, respectively. Thus, from
\begin{equation}
\left(\begin{array}{rr}
-\kappa_2^2 & \omega_3\\
-\omega_2 & -\kappa_3^1\\
\end{array}\right)
=
\frac{t}{2}\left(\begin{array}{cr}
0 & -1 \\
1 & 0\\
\end{array}\right)
+
\frac{s}{2}\left(\begin{array}{cc}
1 & 0 \\
0 & 1\\
\end{array}\right)
+
\left(\begin{array}{cr}
\Delta_1 & \Delta_2 \\
\Delta_2 & -\Delta_1\\
\end{array}\right),
\end{equation}
we can write
\begin{equation}\label{eq::DefModesAsFuncOfkijAndwi}
    t = -(\omega_2+\omega_3),\,s=-(\kappa_2^2+\kappa_3^1), \Delta_1 = \displaystyle\frac{\kappa_3^1-\kappa_2^2}{2}\mbox{ and }
\Delta_2 = \displaystyle\frac{\omega_3-\omega_2}{2}.
\end{equation}
By inverting these relations, we can finally rewrite $\eta_1^2$ and $\eta_1^3$ in equation \eref{eq::omegaijInTermsOfKappaijAndWi} as
\begin{equation}
    \eta_1^2 = -b_{\perp}\eta^1+\left(\frac{s}{2}+\Delta_1\right)\eta^2+\left(-\frac{t}{2}+\Delta_2\right)\eta^3
\end{equation}
and
\begin{equation}
    \eta_1^3 = -b_{\times}\eta^1+\left(\frac{t}{2}+\Delta_2\right)\eta^2+\left(\frac{s}{2}-\Delta_1\right)\eta^3,
\end{equation}
where we write the bend vector as $\mathbf{b}=b_{\perp}\mathbf{n}_2+b_{\times}\mathbf{n}_3$.

The compatibility conditions then come from the structure equations 
\[
\Omega_1^2=\rmd\eta_1^2-\eta_1^3\wedge\eta_3^2, \, \Omega_1^3=\rmd\eta_1^3-\eta_1^2\wedge\eta_2^3,\, \mbox{ and }\, \Omega_2^3=\rmd\eta_2^3-\eta_2^1\wedge\eta_1^3,
\]
where $\Omega_i^j$ are the curvature forms whose coordinates in the basis $\{\eta^i\wedge\eta^j\}$ provide the coefficients of the curvature tensor as defined in equation \eref{eq::CurvedStructureEqs}. Together, the three structure equations provide the $3^4=81$ coefficients of the curvature tensor $R_{ijk\ell}$. From $\Omega_i^j=-\Omega_j^i$ and the fact that $\Omega_i^j$ is a differential form, it follows that $R_{ijk\ell}=-R_{jik\ell}=-R_{ij\ell k}$ reducing these to only $9$ independent entries. However, the first Bianchi identity (which is required to further reduce these to only $6$ independent components) cannot be proved directly from the above definition. Proving this identity requires differentiating $\rmd\eta^i=\eta^k\wedge\eta_k^i$ to obtain $\eta^k\wedge\Omega_k^i=0$, from which follows that $R_{ijk\ell}=R_{k\ell ij}$ (see \cite{doCarmo2012MovingFrames} or \cite{Clelland2017}, p. 376). While these relations hold for  connection forms that are obtained from a moving frame, there could be 1-forms $\eta_i^j$ that would fail to satisfy these relations. Such forms could not be the connection forms of a moving frame in \emph{any} Riemannian geometry. Thus, while for any compatible set of moving frames the Riemann curvature tensor contains only six independent entries, requiring the satisfaction of the first Bianchi identity yields three additional non-trivial compatibility conditions resulting in the following nine equations:
\begin{equation}\label{eq::RijklUsingSplayTwistBend}
\left\{
\begin{array}{lcl}
R_{1212} &=& -(\frac{s}{2}+\Delta_1)_{,1}-b_{\perp,2}-\\[5pt]
         & &-b_{\perp}^2-\frac{s^2}{4}+\frac{t^2}{4}-s\Delta_1-(\Delta)^2+2\omega_1\Delta_2+\kappa_2^1b_{\times},\\[5pt]
R_{1213} &=& -(-\frac{t}{2}+\Delta_2)_{,1}-b_{\perp,3}-\\[5pt]
         & &-b_{\perp}b_{\times}-s(-\frac{t}{2}+\Delta_2)-2\omega_1\Delta_1-\kappa_3^2b_{\times},\\[5pt]
R_{1223} &=& -(-\frac{t}{2}+\Delta_2)_{,2}+(\frac{s}{2}+\Delta_1)_{,3}+\\[5pt]
         & &+tb_{\perp}-2\kappa_2^1\Delta_1+2\kappa_3^2\Delta_2,\\[5pt]
R_{1312} &=& -(\frac{t}{2}+\Delta_2)_{,1}-b_{\times,2}-\\[5pt]
         & &-b_{\perp}b_{\times}-s(\frac{t}{2}+\Delta_2)-2\omega_1\Delta_1-\kappa_2^1b_{\perp},\\[5pt]
R_{1313} &=& -(\frac{s}{2}-\Delta_1)_{,1}-b_{\times,3}-\\[5pt]
         & &-b_{\times}^2-\frac{s^2}{4}+\frac{t^2}{4}+s\Delta_1-(\Delta)^2-2\omega_1\Delta_2+\kappa_3^2b_{\perp},\\[5pt]
R_{1323} &=& -(\frac{s}{2}-\Delta_1)_{,2}+(\frac{t}{2}+\Delta_2)_{,3}+\\[5pt]
         & &+tb_{\times}-2\kappa_2^1\Delta_2-2\kappa_3^2\Delta_1,\\[5pt]
R_{2312} &=& -\kappa_{2,1}^1+\omega_{1,2}-(b_{\times}+\kappa_2^1)(\frac{s}{2}+\Delta_1)\\[5pt]
         & &+(b_{\perp}+\kappa_3^2)(\frac{t}{2}+\Delta_2)+b_{\perp}\omega_1-\kappa_3^2\omega_1,\\[5pt]
R_{2313} &=& \kappa_{3,1}^2+\omega_{1,3}+(b_{\perp}+\kappa_3^2){(\frac{s}{2}-\Delta_1)}-\\[5pt]
         & &-(b_{\times}+\kappa_2^1)(-\frac{t}{2}+\Delta_2)+b_{\times}\omega_1-\kappa_2^1\omega_1,\\[5pt]
R_{2323} &=& \kappa_{3,2}^2+\kappa_{2,3}^1-\\[5pt]
         & &-(\kappa_2^1)^2-(\kappa_3^2)^2-t\,\omega_1-\frac{s^2}{4}-\frac{t^2}{4}+(\Delta)^2,
\end{array}
\right.
\end{equation}
where $f_{,i}=\mathbf{n}_i\cdot\nabla f$ denotes the derivative of $f$ in the direction of $\mathbf{n}_i$ and we denote $\Delta=\sqrt{(\Delta_1)^2+(\Delta_2)^2}$. 

The gradient of the director field can be written in terms of the deformations modes $b_{\perp},b_{\times},s,t,\Delta_1$, and $\Delta_2$. However, notice that by choosing $\mathbf{n}_2$ to be either the normalized bend vector or an eigenvector of the biaxial splay implies we have a Gauge freedom allowing us to set either $b_{\times}=0$ or $\Delta_2=0$. Therefore, this reduces the number of degrees of freedom from 6 to 5. In addition, the equations for the curvature tensor were written in terms of 9 functions, six of which can be written in terms of the deformation modes. Thus, the remaining three, $\kappa_2^1$, $\kappa_3^2$, and $\omega_1$ must be superfluous. We will prove this last assertion in the next two subsections, where we divide the study into director fields with either $\Delta^2=(\Delta_1)^2+(\Delta_2)^2>0$  or $\Delta^2=(\Delta_1)^2+(\Delta_2)^2=0$ on all points. In short, we will have six 6 compatibility equations in 5 functions.

\subsection{Director fields with non-vanishing biaxial splay}

Let us assume non-vanishing biaxial splay $(\Delta)^2=(\Delta_1)^2+(\Delta_2)^2>0$. Then, using the equations for $R_{ijk\ell}$ we can compute the sum $\Delta_1R_{1223}+\Delta_2R_{1323}$, which allows us to  write $\kappa_2^1$ as
\begin{eqnarray}
    \kappa_2^1 & = & -\frac{\Delta_1R_{1223}+\Delta_2R_{1323}}{2\Delta^2}+\frac{\Delta_1t_{,2}-\Delta_2s_{,2}+\Delta_1s_{,3}+\Delta_2t_{,3}}{2\Delta^2}-\nonumber\\
    &-&\frac{\Delta_1\Delta_{2,2}-\Delta_2\Delta_{1,2}}{2\Delta^2}+\frac{(\Delta^2)_{,3}}{4\Delta^2}+t\,\frac{b_{\perp}\Delta_1+b_{\times}\Delta_2}{2\Delta^2}.\label{eq::k21AsFunctionOfDeformationModes}
\end{eqnarray}
Using the equations for $R_{ijk\ell}$ we can find $\Delta_2R_{1223}-\Delta_1R_{1323}$, 
which allows us to write $\kappa_3^2$ as
\begin{eqnarray}
    \kappa_3^2 & = & -\frac{\Delta_1R_{1323}-\Delta_2R_{1223}}{2\Delta^2}-\frac{\Delta_2t_{,2}+\Delta_1s_{,2}-\Delta_1t_{,3}+\Delta_2s_{,3}}{2\Delta^2}+\nonumber\\
    &+&\frac{\Delta_1\Delta_{2,3}-\Delta_2\Delta_{1,3}}{2\Delta^2}+\frac{(\Delta^2)_{,2}}{4\Delta^2}-t\,\frac{b_{\perp}\Delta_2-b_{\times}\Delta_1}{2\Delta^2}.\label{eq::k32AsFunctionOfDeformationModes}
\end{eqnarray}
Analogously, computing $-\Delta_2R_{1212}+\Delta_2R_{1313}+\Delta_1R_{1213}+\Delta_1R_{1312}$ allows us to write $\omega_1$ as
\begin{eqnarray}
    \omega_1 & = & \frac{\Delta_2R_{1212}-\Delta_2R_{1313}-\Delta_1R_{1213}-\Delta_1R_{1312}}{4\Delta^2}+\frac{\Delta_2\Delta_{1,1}-\Delta_1\Delta_{2,1}}{2\Delta^2}-\nonumber\\
    &-&\frac{\Delta_1b_{\times,2}+\Delta_1b_{\perp,3}-\Delta_2b_{\perp,2}+\Delta_2b_{\times,3}}{4\Delta^2}+\frac{(b_{\perp}^2-b_{\times}^2)\Delta_2-2b_{\perp}b_{\times}\Delta_1}{4\Delta^2}-\nonumber\\
    &-&\kappa_2^1\,\frac{b_{\perp}\Delta_1+b_{\times}\Delta_2}{4\Delta^2}+\kappa_3^2\,\frac{b_{\perp}\Delta_2-b_{\times}\Delta_1}{4\Delta^2}.
\end{eqnarray}
Now, substituting the expressions for $\kappa_2^1$ and $\kappa_3^2$ in the equation above, we finally have
\begin{eqnarray}
    \omega_1 & = & \frac{\Delta_2R_{1212}-\Delta_2R_{1313}-\Delta_1R_{1213}-\Delta_1R_{1312}}{4\Delta^2}+\frac{b_{\perp}R_{1223}+b_{\times}R_{1323}}{8\Delta^2}+\nonumber\\
    &+&\frac{\Delta_2\Delta_{1,1}-\Delta_1\Delta_{2,1}}{2\Delta^2}-\frac{\Delta_1b_{\times,2}+\Delta_1b_{\perp,3}-\Delta_2b_{\perp,2}+\Delta_2b_{\times,3}}{4\Delta^2}-\frac{tb^2}{8\Delta^2}-\nonumber\\
    &-&\frac{b_{\times}\Delta_{1,2}-b_{\perp}\Delta_{2,2}+b_{\perp}\Delta_{1,3}+b_{\times}\Delta_{2,3}}{8\Delta^2}+\frac{(b_{\perp}^2-b_{\times}^2)\Delta_2-2b_{\perp}b_{\times}\Delta_1}{4\Delta^2}.\label{eq::w1AsFunctionOfDeformationModes}
\end{eqnarray}

There are three other linearly independent combinations we can construct with the equations for $R_{12ij}$ and $R_{13ij}$. Indeed, using the equations for $R_{ijk\ell}$ we can compute $R_{1212}+R_{1313}$, $R_{1312}-R_{1213}=0$, and $\Delta_1R_{1212}+\Delta_2R_{1213}+\Delta_2R_{1312}-\Delta_1R_{1313}$, which give
\begin{equation}\label{eq::R1212plusR1313AsFunctDefModes}
    R_{1212}+R_{1313} = -s_{,1}-b_{\perp,2}-b_{\times,3}-b^2-\frac{s^2}{2}+\frac{t^2}{2}-2\Delta^2+\kappa_2^1b_{\times}+\kappa_3^2b_{\perp},
\end{equation}
\begin{equation}\label{eq::R1312minusR1213AsFunctDefModes}
    0 =R_{1312}-R_{1213}= -t_{,1}-b_{\times,2}+b_{\perp,3}-st-\kappa_2^1b_{\perp}+\kappa_3^2b_{\times},
\end{equation}
and
\begin{eqnarray}
    s &=& \frac{\Delta_1[R_{1313}-R_{1212}]-\Delta_2[R_{1213}+R_{1312}]}{2\Delta^2}-\nonumber\\
    &-&\frac{[(b_{\perp}^2-b_{\times}^2)\Delta_1+2b_{\perp}b_{\times}\Delta_2]}{2\Delta^2}-\frac{\Delta_1[b_{\perp,2}-b_{\times,3}]+\Delta_2[b_{\times,2}+b_{\perp,3}]}{2\Delta^2}-\nonumber\\
    &-&\kappa_2^1\frac{b_{\perp}\Delta_2-b_{\times}\Delta_1}{2\Delta^2}-\kappa_3^2\frac{b_{\perp}\Delta_1+b_{\times}\Delta_2}{2\Delta^2}.\nonumber\\ \label{eq::sAsFunctDefModes}
\end{eqnarray}
Note we can set $0 =R_{1312}-R_{1213}$ since this is required by the symmetries of the curvature tensor $R_{ijk\ell}$.

Now, substituting $\kappa_2^1$, $\kappa_3^2$, and $\omega_1$ from equations \eref{eq::k21AsFunctionOfDeformationModes}, \eref{eq::k32AsFunctionOfDeformationModes}, and \eref{eq::w1AsFunctionOfDeformationModes} in the three equations we just obtained, we obtain three differential equations of first order involving the deformations modes. In addition, if we also substitute $\kappa_2^1$, $\kappa_3^2$, and $\omega_1$ in the equations for $R_{23ij}$, we will obtain another set of three differential equations involving the deformations modes and their first and second derivatives. 

\subsection{Director fields with vanishing biaxial splay}

Now, let us assume a vanishing biaxial splay $(\Delta)^2=(\Delta_1)^2+(\Delta_2)^2=0$. If we  assume that $\mathbf{b}\not=0$, then from the equations of $R_{ijk\ell}$ we can compute $b_{\times}R_{1212}-b_{\perp}R_{1312}$, which allows us to write $\kappa_2^1$ as
\begin{eqnarray}
    \kappa_2^1 &=& \frac{b_{\times}R_{1212}-b_{\perp}R_{1312}}{b^2}+\frac{b_{\times}s_{,1}-b_{\perp}t_{,1}}{2b^2}+\frac{b_{\perp}b_{\times,2}-b_{\times}b_{\perp,2}}{b^2}-\nonumber\\
    &-&\frac{(t^2-s^2)b_{\times}+2stb_{\perp}}{4b^2}.\label{eq::k21AsFunctDefModesDeltaZero}
\end{eqnarray}
In addition, from the equations of $R_{ijk\ell}$ we can compute  $b_{\perp}R_{1313}-b_{\times}R_{1213}$, which allows us to write $\kappa_3^2$ as
\begin{eqnarray}
    \kappa_3^2 &=& \frac{b_{\perp}R_{1313}-b_{\times}R_{1213}}{b^2}+\frac{b_{\times}t_{,1}+b_{\perp}s_{,1}}{2b^2}+\frac{b_{\perp}b_{\times,3}-b_{\times}b_{\perp,3}}{b^2}+\nonumber\\
    &+&\frac{(s^2-t^2)b_{\perp}+2stb_{\times}}{4b^2}.\label{eq::k32AsFunctDefModesDeltaZero}
\end{eqnarray}
Note that when $\Delta=0$, we can write $\omega_1$ as a function of the deformation modes $\{s,t,b_{\perp},b_{\times}\}$ by substituting for $\kappa_2^1$ and $\kappa_3^2$ in the equation for $R_{2323}$. Alternatively, we can get rid of $\omega_1$ by choosing $\mathbf{n}_2$ and $\mathbf{n}_3$ such that $\omega_1=0$.

On the other hand, if $\mathbf{b}=0$, then there are some restrictions on the geometry of the ambient manifold. Indeed, we straightforwardly conclude that $R_{1212}=R_{1313}$ and $R_{1213}=-R_{1312}$, which by using the symmetry $R_{1213}=R_{1312}$ allows us to deduce that $R_{1213}=R_{1312}=0$.

\section{Uniform distortion director fields on manifolds of constant curvature}

In this section, we provide a characterization of uniform distortion fields, i.e., director fields $\mathbf{n}$ for which the deformation modes $\{s,t,b,\Delta_1,\Delta_2\}$ are all constant, in manifolds of constant sectional curvature. As a consequence, it will follow that no combination of values other than the pure twist phase exist in positive curvature. For negative curvature, the examples of uniform distortion fields with $s^2+4b^2=4$ and $t=0$ provided in \cite{SadocNJP2020} are in fact the most general case under the assumption that the biaxial splay vanishes. However, our results will also imply that it is possible to have uniform distortion fields in negative curvature with non-vanishing biaxial splay and, as in the Euclidean space, these phases correspond to foliations of space by helices. 

From now on, let us assume that we have a director field in a curved space $M^3$ of constant curvature $R_0$. This means that the curvature tensor is given by $R_{ijk\ell}=R_0(\delta_{ik}\delta_{j\ell}-\delta_{i\ell}\delta_{jk})$ \cite{Spivak4}: $R_0<0$ if $M^3$ is locally isometric to a hyperbolic space, $R_0=0$ if $M^3$ is locally isometric to the Euclidean space, and $R_0>0$ if $M^3$ is locally isometric to a {three-sphere}. Let us also introduce the shorthand notation $\langle D\mathbf{b},\mathbf{b}\rangle=(b_{\perp}^2-b_{\times}^2)\Delta_1+2b_{\perp}b_{\times}\Delta_2=b^2\Delta \cos(2\phi)$ and $\langle JD\mathbf{b},\mathbf{b}\rangle=(b_{\times}^2-b_{\perp}^2)\Delta_2+2b_{\perp}b_{\times}\Delta_1=b^2\Delta \sin(2\phi)$, where $\phi$ is the angle formed by the bend vector and the principal direction of the biaxial splay. (In other words, $D$ and $J$ denote the biaxial splay and the counterclockwise $\frac{\pi}{2}$-rotation acting as linear operators on the plane normal to the director field, respectively.) The results of this section can be summarized as follows
\begin{theorem}
Let $M^3$ be a manifold of constant sectional curvature $R_0$ and let $\mathbf{n}$ be a director field in it with constant deformation modes $\{s,t,b_{\perp},b_{\times},\Delta_1,\Delta_2\}$.
\begin{enumerate}[(a)]
    \item If $R_0>0$, then $s=b=\Delta=0$ and $t=\pm2\sqrt{R_0}$ is the only solution. 
    \item If $R_0=0$, then $b=s=t=0$ when $\Delta=0$. On the other hand, when $\Delta\not=0$, then $s=0$, $t=\pm2\Delta$, and $\phi=\frac{(2k+1)\pi}{4}$, $k\in\{0,1,2,3\}$, i.e., the bend vector $\mathbf{b}$ bisects the principal directions of the biaxial splay, where $k=0$ or $k=2$ if $t=2\Delta$ and $k=1,3$ if $t=-2\Delta$. 
    \item If $R_0<0$, then $t=0$ and $s^2+4b^2=-4R_0$ when $\Delta=0$. On the other hand, when $\Delta\not=0$, then $t=\pm2\Delta$, $s=-\frac{1}{2\Delta^2}\langle D\mathbf{b},\mathbf{b}\rangle$, and the deformation modes are subjected to the restriction $1+\frac{b^2}{4\Delta^2}\geq\sqrt{-\frac{R_0}{\Delta^2}}$.
\end{enumerate}
In (b) and (c), the bend and biaxial splay are the free parameters describing the families of solutions.
\end{theorem}

\begin{figure}[t]
    \centering
    \includegraphics[width=0.75\linewidth]{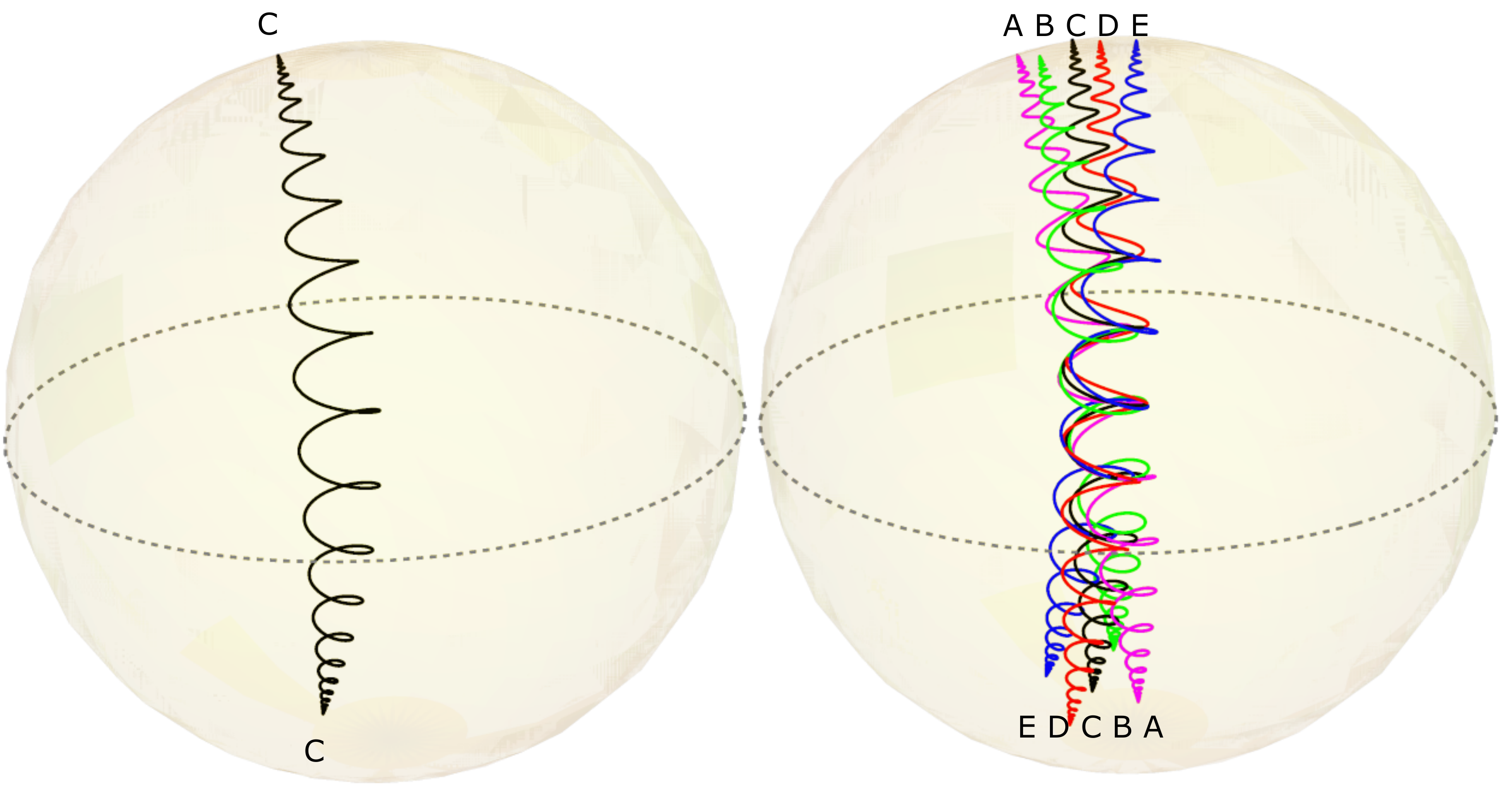}
    \caption{{(left) A single integral curve of a uniform distortion field $\mathbf{n}$ in hyperbolic space $\mathbb{H}^3$ using the Poincar\'e ball model. (right) A collection of five integral curves of a uniform distortion field in hyperbolic space $\mathbb{H}^3$ forming congruent helices of similar curvature $\kappa=b$ and torsion $\tau=\omega_1$. The collection of all integral curves constitute a foliation of $\mathbb{H}^3$ by congruent helices.  
    The dashed gray line indicates the points $(x,y,z)$ in the ideal boundary with $z=0$. The different helices are assigned five distinct letters (colors) to help guide the eye. Note that the distance between the integral curves increases as we move away from the point $(0,0,0)$ indicating a non-zero splay. This is expected in hyperbolic space since the failure of the Parallel Postulate is equivalent to the non-existence of equidistant geodesics and, therefore, we expect to  not have equidistant congruent curves in $\mathbb{H}^3$.  (In the figure, the sectional curvature is $R_0=-1$, $b = 4.0$ [with $b_{\perp}=-b$ and $b_{\times}=0$], $\Delta_1 = \Delta_2 = \frac{\sqrt{2}}{2}$, $t=2\Delta=2$, and $s=-\frac{1}{2\Delta^2}[(b_{\perp}^2-b_{\times}^2)\Delta_1+2b_{\perp}b_{\times}\Delta_2]=-4\sqrt{2}$.)}}
    \label{fig:H3UniDistField}
\end{figure}

{Illustrations of uniform distortion fields in the Euclidean space and 3-sphere can be found in Refs. \cite{VirgaPRE2019} and \cite{SadocNJP2020}, respectively. In our Figure \ref{fig:H3UniDistField} we illustrate a uniform distortion field in hyperbolic space. As we show below, similarly to the case of Euclidean space, in $\mathbb{H}^3$ uniform distortions also give rise to foliation of space by congruent helices.} 

In the next subsections we are going to provide a proof for this theorem by analyzing the restrictions imposed by the compatibility equations on the values of the deformation modes. But, before that, let us discuss the implications on the geometry of the integral curves of the director field. 

For $R_0>0$, it is known that the vector field tangent to the fibers of the Hopf fibration provides an example of a uniform distortion field \cite{SadocNJP2020,SethnaPRL1983}. It turns out that this is the only possibility. Indeed, given any uniform distortion field $\mathbf{n}$ on a manifold of constant positive curvature, the integral curves of $\mathbf{n}$  are geodesics. In addition, from the fact that $s,\Delta_1$, and $\Delta_2$ all vanish, we deduce that any two integral curves are parallel to each other, from which follows that the fibration provided by the integral curves of $\mathbf{n}$ is locally a Hopf fibration \cite{NuchiAGT2015}. 

For $R_0\leq0$, the integral curves of a uniform distortion field do not have to be  geodesics. In general, they are helices, i.e., curves with constant curvature and torsion. Indeed, the equations of motion of the  $\{\mathbf{n}_1,\mathbf{n}_2,\mathbf{n}_3\}$ are 
\begin{equation*}
\nabla_{\mathbf{n}_1}\left(
\begin{array}{c}
\mathbf{n}_1\\
\mathbf{n}_2\\
\mathbf{n}_3\\
\end{array}\right)
=\left(
\begin{array}{ccc}
0 & -b_{\perp} & -b_{\times}\\[4pt]
b_{\perp} & 0 & \omega_1\\[4pt]
b_{\times} & -\omega_1 & 0\\
\end{array}\right)\left(
\begin{array}{c}
\mathbf{n}_1\\
\mathbf{n}_{2}\\
\mathbf{n}_{3}\\
\end{array}\right),
\end{equation*}
where $\omega_1$ is constant and given by equation \eref{eq::w1AsFunctionOfDeformationModes}. We can obtain the Frenet frame $\{\mathbf{T}=\mathbf{n},\mathbf{N},\mathbf{B}\}$ from $\{\mathbf{n}_1,\mathbf{n}_2,\mathbf{n}_3\}$ by a rotation of an angle $\theta$ on the normal plane. Then, we can write $\kappa_1^1=\kappa\cos\theta$, $\kappa_1^2=\kappa\sin\theta$, and $\theta'=\tau-\omega_1$ \cite{TakagiPTP1992}. Since $\kappa_1^i$ and $\omega_1$ are all constant, we deduce that $\kappa$ and $\tau$ are also constant. As a consequence, the integral curves of $\mathbf{n}$ form helices and make a constant angle with the Darboux vector field $\mathbf{w}=\omega_1\mathbf{n}_1-\kappa_1^2\mathbf{n}_2+\kappa_1^1\mathbf{n}_3$ or $\mathbf{w}=\tau \mathbf{T}+\kappa \mathbf{B}$ if we use the Frenet frame:  if $s$ denotes the arc-length of the integral curves of $\mathbf{n}$, then $\frac{\rmd}{\rmd s}\langle\mathbf{w},\mathbf{n}\rangle=\langle\nabla_{\mathbf{n}}\mathbf{w},\mathbf{n}\rangle+\langle\mathbf{w},\nabla_{\mathbf{n}}\mathbf{n}\rangle=\langle \tau'\mathbf{n}+b'\mathbf{B}+\tau b\mathbf{N}-\tau b \mathbf{N},\mathbf{n}\rangle+\langle\mathbf{w},b\mathbf{N}\rangle=0$.  

Virga's strategy to characterize the helicoidal phases in Euclidean space \cite{VirgaPRE2019} consisted in investigating the behavior of the frame along a generic curve in space (not necessarily an integral curve). This gives rise to an operator whose eigenvector can be shown to be constant and, in addition, the integral lines of the director field precess around this fixed direction. We can provide an alternative proof by showing that all integral curves of a uniform distortion field $\mathbf{n}$ in a flat manifold have the same axis, i.e., we may show that $\mathbf{v}\cdot\nabla\mathbf{w}=0$ for every direction $\mathbf{v}$.

Using equations \eref{eq::EqsOfMotionMovFrame} and  \eref{eq::DefModesAsFuncOfkijAndwi}  to write some of the $\kappa_i^j$'s and $\omega_i$'s as functions of the deformation modes, we conclude that
\begin{equation*}
    \mathbf{w}_{,1} = (b_{\times}b_{\perp}-b_{\perp}b_{\times})\mathbf{n}_1+(b_{\perp}\omega_1-\omega_1b_{\perp})\mathbf{n}_2+(b_{\times}\omega_1-\omega_1b_{\times})\mathbf{n}_3=0,
\end{equation*}
\begin{eqnarray*}
    \mathbf{w}_{,2} &=& \left[b_{\perp}\left(\frac{t}{2}+\Delta_2\right)-b_{\times}\Delta_1\right]\mathbf{n}_1+(\omega_1\Delta_1+b_{\perp}\kappa_2^1)\mathbf{n}_2+\\
    &+&\left[\omega_1\left(\frac{t}{2}+\Delta_2\right)+b_{\times}\kappa_2^1\right]\mathbf{n}_3,
\end{eqnarray*}
and
\begin{eqnarray*}
    \mathbf{w}_{,3} &=& \left[b_{\times}\left(\frac{t}{2}-\Delta_2\right)+b_{\perp}\left(\frac{s}{2}-\Delta_1\right)\right]\mathbf{n}_1-\left[\omega_1\left(\frac{t}{2}-\Delta_2\right)+b_{\perp}\kappa_3^2\right]\mathbf{n}_2+\\
    &+&\left[\omega_1\left(\frac{s}{2}-\Delta_1\right)-b_{\times}\kappa_3^2\right]\mathbf{n}_3.
\end{eqnarray*}

\begin{figure}[t]
    \centering
    \includegraphics[width=\linewidth]{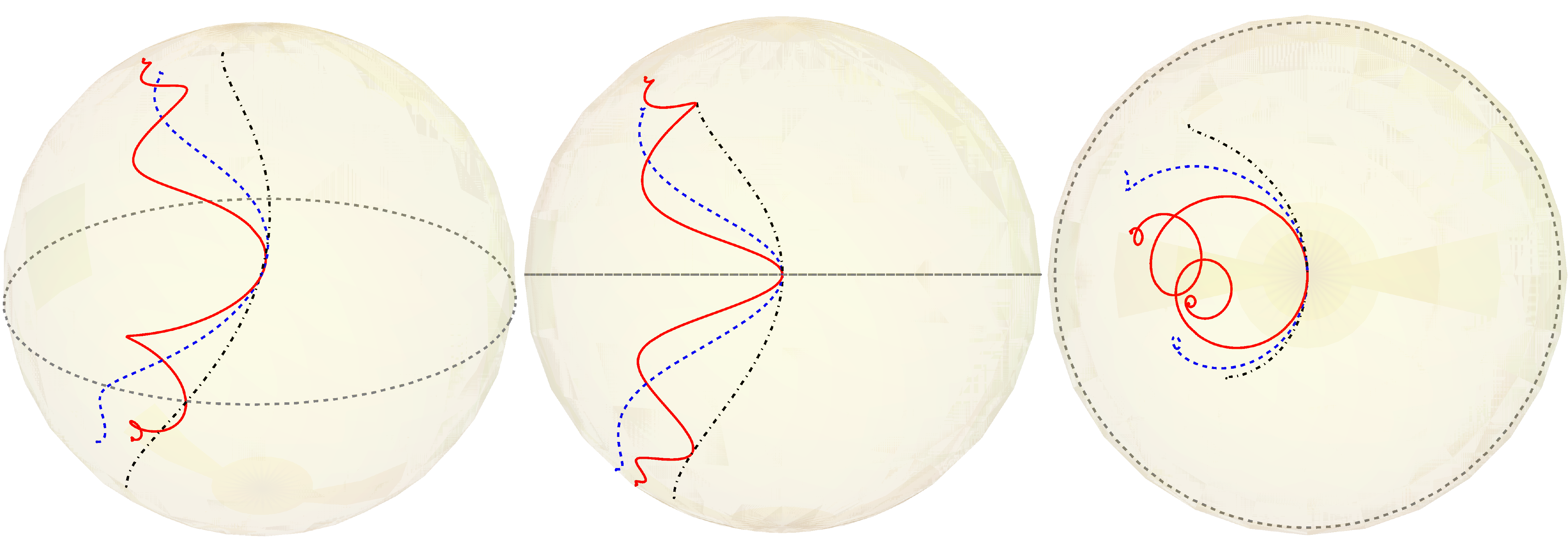}
    \caption{{
    Helices in $\mathbb{H}^3$ in the Poincar\'e ball model shown, from left to right in perspective-view, side-view ($xz$-plane projection), and top-view ($xy$-plane projection), respectively.  The dashed gray line indicates the points $(x,y,z)$ in the ideal boundary with $z=0$. Up to rigid motions in $\mathbb{H}^3$, every hyperbolic helix $\gamma$ is parametrized by $\gamma(t)=\frac{1}{1+\cosh(\theta)\cosh(\omega_1t)}(\cosh(\theta)\sinh(\omega_1t),\sinh(\theta)\cos(\omega_2t),\sinh(\theta)\sin(\omega_2t))$, where $\theta=\theta(\omega_1,\omega_2)$ and $\omega_1,\omega_2$ are constants depending on $\kappa$ and $\tau$. In the figure, the sectional curvature is $R_0=-1$, the torsion is $\tau=\frac{1}{2}$ while the curvature is $\kappa=\frac{1}{2}$ for the dot-dashed black curve, $\kappa=1$ for the dashed blue curve, and $\kappa=\frac{3}{2}$ for the full red curve. Note that the oscillatory behavior of $\gamma$ is more prominent for large values of $\kappa$.}}
    \label{fig:H3helices}
\end{figure}

On the one hand, for $R_0=0$, substitution of the values of the deformation modes of a uniform distortion field allows us to deduce that $\mathbf{n}_i\cdot\nabla\mathbf{w}=0$, $i=1,2,3$. Therefore, $\mathbf{n}$ provides a foliation of space by parallel helices. On the other hand, for $R_0<0$, we still have that $ \mathbf{w}_{,1}=0$ and, therefore, $\mathbf{w}$ is parallel transported along the integral curves of $\mathbf{n}$. On the other hand, in general $ \mathbf{w}_{,2}$ and $ \mathbf{w}_{,3}$ do not vanish, implying that $\mathbf{n}$ provides a foliation of hyperbolic space by helices which are not necessarily parallel. In a hyperbolic space, we need to distinguish between three types of helices. First notice that a curve with zero torsion is necessarily contained in a totally geodesic surface, i.e., locally the surface is a copy of a hyperbolic plane of curvature $R_0$. There are three types of planes curves with constant curvature $b>0$: circles if $b\in(\sqrt{-R_0},\infty)$, horocycles if $b=\sqrt{-R_0}$, and hypercycles if $b\in(0,\sqrt{-R_0})$ \cite{Ramsay1994}. Therefore, depending on the values of the bend $b$, we expect three families of helices in hyperbolic geometry. {Representative members of each of the families of hyperbolic helices are illustrated in Figure \ref{fig:H3helices}.}

\subsection{Uniform distortion fields with vanishing biaxial splay $(\Delta)^2=(\Delta_1)^2+(\Delta_2)^2=0$}

First, assume we have $b_{\perp}=b_{\times}=0$. Then, from the equations for $R_{1212}$ and $R_{1213}$, it follows that $R_0=\frac{1}{4}(t^2-s^2)$ and $st =0$. Consequently, $s=0$ or $t=0$ and we finally conclude
\begin{equation}
\Delta=0,\,b=0\Rightarrow\left\{
\begin{array}{lcc}
s = 0\mbox{ and }t=\pm2\sqrt{R_0} & \mbox{ if } & R_0\geq0\\
t = 0\mbox{ and }s=\pm2\sqrt{-R_0} & \mbox{ if } & R_0\leq0\\
\end{array}
\right..
\end{equation}
In particular, in Euclidean space, $\Delta=0$ and $b=0$ imply that the director field is constant: $\rmd\mathbf{n}\equiv0$.

Now, assume that $\Delta=0$ but $b\not=0$. From $R_{1223}=0$ and $R_{1323}=0$, we necessarily have $t=0$. From equations \eref{eq::k21AsFunctDefModesDeltaZero} and \eref{eq::k32AsFunctDefModesDeltaZero} it follows
\begin{equation}
    \kappa_2^1 = \frac{b_{\times}}{b^2}\left(R_0+\frac{s^2}{4}\right)\mbox{ and }
    \kappa_3^2 = \frac{b_{\perp}}{b^2}\left(R_0+\frac{s^2}{4}\right).
\end{equation}
Substituting the expressions for $\kappa_2^1$, $\kappa_3^2$ in the equation for $R_{2323}$ gives
\begin{equation}
    R_0 = -\frac{s^2}{4}-(R_0+\frac{s^2}{4})^2\frac{b_{\perp}^2+b_{\times}^2}{b^4} \Rightarrow  \frac{1}{b^2}\left(R_0+\frac{s^2}{4}\right)\left(b^2+R_0+\frac{s^2}{4}\right)=0.
\end{equation}
Therefore, $s^2=-4R_0$ or $s^2=-4b^2-4R_0$. 

On the one hand, we see that if $R_0\geq0$, then there exists no solution with $b\not=0$. On the other hand, if $R_0<0$, we could equally have either  $s^2=-4R_0$ or $s^2=-4b^2-4R_0$ (there is no sign obstruction for $R_0<0$). However, only $s^2=-4b^2-4R_0$ is allowed. Indeed, if it were $s^2=-R_0$, then substituting the expressions for $\kappa_2^1$ and $\kappa_3^2$ above in the equations for $R_{1212}$, $R_{1313}$ and summing them would give
\begin{equation}
    2R_0 = -b^2-\frac{s^2}{2}+R_0+\frac{s^2}{4}\Rightarrow R_0 = -b^2 -\frac{s^2}{4}=-b^2-\frac{(-4R_0)}{4}\Rightarrow b^2 = 0.
\end{equation}
This contradicts the assumption that $b\not=0$. Finally, we conclude that
\begin{equation}
\Delta=0,\,b\not=0\Rightarrow\left\{
\begin{array}{ccc}
\nexists\mbox{ solution} & \mbox{ if } & R_0>0\\
t = 0\mbox{ and }s^2+4b^2=-4R_0 & \mbox{ if } & R_0\leq0\\
\end{array} 
\right..
\end{equation}
Notice that in the case $R_0<0$, such as in hyperbolic space, the configuration with $\Delta=0$ and $t=0$ becomes the trivial director field in Euclidean space in the limit $R_0\to0^{-}$.

\subsection{Uniform distortion fields with non-vanishing biaxial splay $(\Delta)^2=(\Delta_1)^2+(\Delta_2)^2>0$}

If all deformation modes are constant, it follows from equations \eref{eq::k21AsFunctionOfDeformationModes}, \eref{eq::k32AsFunctionOfDeformationModes}, and \eref{eq::w1AsFunctionOfDeformationModes} that $\kappa_2^1$, $\kappa_3^2$, and $\omega_1$ are also constant and equal to
\begin{equation}
    \kappa_2^1 = t\,\frac{b_{\perp}\Delta_1+b_{\times}\Delta_2}{2\Delta^2},\kappa_3^2=-t\,\frac{b_{\perp}\Delta_2-b_{\times}\Delta_1}{2\Delta^2},\omega_1 = -\frac{tb^2}{8\Delta^2}-\frac{\langle JD\mathbf{b},\mathbf{b}\rangle}{4\Delta^2}.
\end{equation}

Now, substituting $\kappa_2^1$ and $\kappa_3^2$ in equation \eref{eq::sAsFunctDefModes}, implies that the splay is given by
\begin{equation}\label{eq::sAsFunctDefModesSpaceForm}
    s =  -\frac{\langle D\mathbf{b},\mathbf{b}\rangle}{2\Delta^2}.
\end{equation}
In addition, substituting $\kappa_2^1$, $\kappa_3^2$, and $\omega_1$ in equations \eref{eq::R1212plusR1313AsFunctDefModes}, \eref{eq::R1312minusR1213AsFunctDefModes}, and $R_{2323}$  from equation \eref{eq::RijklUsingSplayTwistBend}, gives
\begin{equation}\label{eq::R1212plusR1313AsFunctDefModesSpaceForm}
    2R_{0} = -b^2-\frac{s^2}{2}+\frac{t^2}{2}-2\Delta^2+t\frac{(b_{\times}^2-b_{\perp}^2)\Delta_2+2b_{\perp}b_{\times}\Delta_1}{2\Delta^2},
\end{equation}
\begin{equation}\label{eq::R1312minusR1213AsFunctDefModesSpaceForm}
    0 = -st-t\frac{(b_{\perp}^2-b_{\times}^2)\Delta_1+2b_{\perp}b_{\times}\Delta_2}{2\Delta^2},
\end{equation}
and
\begin{equation}
     R_0 =  -\frac{s^2}{4}-\frac{t^2}{4}+\Delta^2-\frac{t^2b^2}{8\Delta^2}+t\frac{(b_{\times}^2-b_{\perp}^2)\Delta_2+2b_{\perp}b_{\times}\Delta_1}{4\Delta^2}.
\end{equation}
Notice that from the expression we got for the splay in equation \eref{eq::sAsFunctDefModesSpaceForm}, it follows that equation \eref{eq::R1312minusR1213AsFunctDefModesSpaceForm} is redundant. Subtracting equation \eref{eq::R1212plusR1313AsFunctDefModesSpaceForm} from twice the last equation above allows us to conclude that
\begin{equation}
    0 = \left(1+\frac{b^2}{4\Delta^2}\right)(t^2-4\Delta^2)\Rightarrow t = \pm2\Delta.
\end{equation}

\begin{remark}
The equations for $R_{2312}$ and $R_{2313}$ in \eref{eq::RijklUsingSplayTwistBend} provide no further constraints. In fact, substituting $\kappa_2^1$, $\kappa_3^2$, and $\omega_1$ in $R_{2312}$ and in $R_{2313}$ respectively gives
\[
     \left(1+\frac{b^2}{4\Delta^2}\right)\left(1-\frac{t^2}{4\Delta^2}\right)\left(b_{\times}\Delta_1-b_{\perp}\Delta_2\right)=0
\]
and 
\[
\left(1+\frac{b^2}{4\Delta^2}\right)\left(1-\frac{t^2}{4\Delta^2}\right)\left(b_{\perp}\Delta_1+b_{\times}\Delta_2\right)=0.
\]
Now, taking into account that $t=\pm2\Delta$, which is obtained from $R_{2323}$, the two equations above vanish identically.
\end{remark}

Let us write $t=2\delta\Delta$, $\delta=\pm1$, and substitute for $t$ and $s$ in equation \eref{eq::R1212plusR1313AsFunctDefModesSpaceForm}. Thus,
\begin{equation}
     2R_0 = -b^2-\frac{\langle D\mathbf{b},\mathbf{b}\rangle^2}{8\Delta^4}+\delta\frac{\langle JD\mathbf{b},\mathbf{b}\rangle}{\Delta},
\end{equation}
from which we find that
\begin{equation}\label{eq::implicitEqForR0DeltaBend}
    2 R_0\Delta+\frac{\langle D\mathbf{b},\mathbf{b}\rangle^2}{8\Delta^3}+b^2\Delta+\delta\langle JD\mathbf{b},\mathbf{b}\rangle=0.
\end{equation}
Now, from the Cauchy-Schwarz inequality, it follows that
$$
\vert \langle JD\mathbf{b},\mathbf{b}\rangle \vert \leq \Vert JD\mathbf{b}\Vert \Vert\mathbf{b}\Vert\leq  b^2\Delta\Rightarrow 0\leq  b^2\Delta+\delta\langle JD\mathbf{b},\mathbf{b}\rangle.
$$
We immediately have the following conclusions:
\begin{enumerate}[(a)]
    \item If $R_0>0$, then equation \eref{eq::implicitEqForR0DeltaBend} is a sum of non-negative numbers. However, $R_0\Delta>0$ and, consequently, there must be no uniform director field with $\Delta>0$ on a space of constant positive curvature, such as the {three-sphere}.
    \item If $R_0=0$, then we must have $\langle D\mathbf{b},\mathbf{b}\rangle=0$ and also  $b^2\Delta+\delta\langle JD\mathbf{b},\mathbf{b}\rangle=0$. It follows that in a space of vanishing curvature the splay $s$ must vanish and the bend vector $\mathbf{b}$ bisects the principal directions of the biaxial splay, i.e., $\phi=\frac{(2k+1)\pi}{4}$, $k\in\{0,1,2,3\}$, where $k=0$ or $k=2$ if $t=2\Delta$ and $k=1,3$ if $t=-2\Delta$.
\end{enumerate}
    
It remains to further analyze uniform distortion director fields in hyperbolic geometry, i.e., $R_0<0$. Seeing equation \eref{eq::R1212plusR1313AsFunctDefModesSpaceForm} as a quadratic polynomial in $t$, its discriminant is
\begin{equation}
     \mbox{disc.} =4\Delta^2\left(1+\frac{b^2}{4\Delta^2}\right)^2+4R_0.
\end{equation}
Thus, the requirement that $t\in\mathbb{R}$ demands $\mbox{disc.}\geq0$, which implies
\begin{equation}
    1+\frac{b^2}{4\Delta^2} \geq \sqrt{-\frac{R_0}{\Delta^2}}.
\end{equation}
Note that if we choose $\Delta\geq\sqrt{-R_0}$, then the above inequality imposes no restriction on the values of the bend $b$.

\section{Discussion}
In the intrinsic approach materials are described only through quantities available to an observer residing within the material \cite{ME21}. These quantities may be associated with some non-trivial locally preferred reference values that manifest the constituents' shape and mutual interactions. We show that a collection of five such scalar (and pseudoscalar) fields suffice to characterize the director texture. These fields can be chosen to be the bend, splay, twist, saddle-splay and the relative orientation between the principal biaxial splay direction and the bend direction. 

In 2D only two such fields suffice to uniquely prescribe a director field. The compatibility conditions in the 2D case amount to a single first order differential relation \cite{NivSM2018}. In three dimensions we obtained six differential relations. Three of first order, and three of second order. Thus the system is of at most second order; it is presently unknown whether the system can be further reduced to yield a purely first order system or not. Understanding the 
degree and structure of the compatibility conditions is important not for taxonomical reasons, but rather as these
determine the super-extensive rate at which energy accumulates when a frustrated phase grows in size \cite{ME21}. 

Though more abstract than an approach uniquely based on vector calculus, the method of moving frames allows us to obtain manageable equations and to investigate  director  fields  in  both  Euclidean  and  curved  Riemannian  spaces in an equal foot. In particular it allows us to find all uniform distortion fields for all isotropic homogeneous Riemannian manifolds. 

{The exhaustive nature of the compatibility conditions presented here allows us to assert that the well known constant twist phase in $\mathbb{S}^3$ is, in fact, the only uniform distortion field $\mathbb{S}^3$ supports. For $\mathbb{H}^3$ we extend the result of Sadoc, Mosseri and Selinger \cite{SadocNJP2020} who found particular solutions with vanishing twist and biaxial splay, to show that these constitute all possible textures foliated by planar curves in $\mathbb{H}^3$. Moreover, we find all the textures foliated by non-planar curves characterized by non-vanishing twist and biaxial splay that grow in proportion to each other. In general, we showed that uniform distortion fields in $\mathbb{H}^3$ yield textures foliated by congruent helices, encompassing the previous results.} 

{
The full compatibility conditions provided here allow us to extend our understanding of three dimensional frustrated textures in Euclidean space to the realm of non-uniform distortion fields. Small enough domains of bent-core liquid crystals are expected to allow a non-uniform distortion field associated with an elastic energy that is lower than that of the uniform twist-bent phase \cite{Vir14}. Knowledge of the full compatibility conditions provides a path for constructing such low energy solutions for small enough domains: starting with a state of pure bend at some point in the domain, satisfaction of the compatibility conditions necessitates certain gradients to assume a non-vanishing value. Incorporating the constitutive law at this point may help select which of these gradients will be chosen to balance the attempted constant pure bend. Such ``propagation" of solutions may also find use in solving the inverse design of three dimensional responsive material, analogously to the procedure carried out for 2D in \cite{GAE19}.
}

Considering the five characterizing fields as given quantities and solving for the corresponding director field may also be carried out as long as the compatibility conditions are satisfied. The compatibility conditions, in turn, can be interpreted both in terms of a {material} frame, where the fields {gradients are given in terms of their projections on the director orientations}, and a lab frame in which the field {gradients} are given explicitly in terms of the embedding space coordinates. It is important to note that the information contained in the two viewpoints is not equivalent. The material intrinsic description, which is more natural, may be used to integrate the director field from knowledge about its local behavior. Such an approach is particularly useful for solving inverse design problems \cite{GAE19}, and for constructing new optimal textures. The {lab frame}  approach is somewhat less natural as it assumes that the fields are given in terms of the stationary embedding space coordinates, yet the director is unknown. For the two dimensional case the lab-frame approach allowed obtaining the director field directly from the gradients of the bend and splay functions, {provided they were compatible \cite{NivSM2018}. For general fields, $b$ and $s$, one may apply the reconstruction formula provided that the gradients of these fields are large enough. However, this does not assure that these fields were indeed compatible. In this approach the compatibility condition is replaced by self-consistency conditions equating the splay and bend of the resulting director field with those used to generate it. One may expect that these conditions will produce two second order differential equations for the fields $b$ and $s$ that are independent of the director orientation, however to the best of our knowledge these relations have not yet been obtained. }

When finishing this manuscript, it came to our attention that a similar approach to the one presented here was recently pursued by Pollard and Alexander \cite{PollardArXiv2021}. There, the authors also develop the idea of using the moving frame method to obtain the compatibility conditions for the deformation modes of a director field expressed in terms of the curvature tensor of the ambient manifold. {In addition, they present the lab frame reconstruction formulae for the director in terms of the fixed frame spatial gradients of the deformation modes. These, much like their two dimensional analogs are assured to satisfy the self-consistency conditions if the scalar deformations modes were indeed compatible, yet for general fields do not have to be self-consistent.} Finally, they discuss the examples constructed by Sadoc \emph{et al} \cite{SadocNJP2020} in the language of moving frames by exploiting the fact that there exists an underlying Lie algebra structure associated with uniform distortion fields. While the choice of applications differs between our work and that of Pollard and Alexander, the main guiding principles and calculations of the components of the Riemann curvature tensor are similar. {In particular the nine equations (19-25) and (31-33) in their manuscript can be directly translated to the nine equations derived here (\ref{eq::RijklUsingSplayTwistBend}). While we further reduce these nine equations in eight unknown fields to six equations in the five deformation modes, the compatibility conditions remain equivalent, and any set of deformation modes determined compatible by one of the methods would be compatible with respect to the other. 
} 

\ack
This work was funded by the Israel Science Foundation grant no. $1479/16$. E.E. thanks the Ascher Foundation for their support. L.C.B.dS. acknowledges the support provided by the Mor\'a Miriam Rozen Gerber fellowship for Brazilian postdocs.

\end{document}